# Molecular dynamics of a short range ordered smectic phase nanoconfined into porous silicon.

R. Guégan[1], D. Morineau[1], R. Lefort[1], A. Moréac[1], W. Béziel[1], M. Guendouz[2], J.-M. Zanotti[3], B. Frick[4]

[1]*Groupe Matière Condensée et Matériaux, CNRS-UMR 6626, Bâtiment 11A, Université de Rennes 1, F-35042 Rennes, France*

[2]*Laboratoire d'Optronique, FOTON, CNRS-UMR 6082, Université de Rennes 1, F-22302 Lannion Cedex, France*

[3]*Laboratoire Léon Brillouin (CEA-CNRS), F-91191 Gif-sur-Yvette Cedex, France*

[4]*Institut Laue-Langevin, 6 rue Jules Horowitz, F-38042 Grenoble, Cedex 9, France*

**ABSTRACT**

4-n-octyl-4-cyanobiphenyl (8CB) has been recently shown to display an unusual sequence of phases when confined into porous silicon (PSi). The gradual increase of oriented short-range smectic (SRS) correlations in place of a phase transition has been interpreted as a consequence of the anisotropic quenched disorder induced by confinement in PSi. Combining two quasielastic neutron scattering experiments with complementary energy resolutions, we present the first investigation of the individual molecular dynamics of this system. A large reduction of the molecular dynamics is observed in the confined liquid phase, as a direct consequence of the dynamical boundary conditions imposed by the confinement. Temperature fixed window scans (FWS) reveal a continuous 'glass-like' reduction of the molecular dynamics of the confined liquid and SRS phases on cooling down to 250 K, where a solid-like behavior is finally reached by a two steps crystallization process.

## 1. Introduction

Molecular fluids confined in porous materials of nanometer size usually exhibit very exceptional features as compared to bulk systems. Major effects are observed when the system size is sub-micrometric i.e. of the order of a few tens of molecular diameters. Reported studies of nanoconfined systems suggest that their unusual behavior may originate from intricate finite size, interfacial, low-dimensionality and/or quenched disorder effects.[1,2,3,4,5]

Most noticeably, confined systems show strongly modified phase behavior with absent or strongly depressed phase transitions, new structural orders and an improved tendency to supercooling and metastability.[6] Thermodynamical quantities such as density or structure factors have been recently shown to be affected too.[7,8,9]

At present, some most challenging issues for fundamental research are related to the dynamics of confined phases. An area of intense activity concerns the structural relaxation processes and the glassy dynamics of confined liquids.[10,11] The motivation is threefold: confinement facilitates supercooling and allows one to study the glass transition of very simple molecules, which is not possible in bulk.[5] Confinement is also expected to tell about cooperative relaxation processes that appear at the nanometric scale in supercooled liquids.[2,10,11] Finally, even normal liquids (i.e. above the melting point) can exhibit dynamical features in confinement, which are commonly related to the phenomenology of the glass transition in terms of non-Debye character of the structural relaxation.[12]

Liquid crystals (LC's) are recognized as especially interesting for various technological applications. The occurrence of translational symmetry and orientational order has been extensively studied in confined geometry.[13] However, many of the physical properties of LC's remain open when the confinement reaches nanometric sizes. At this lengthscale, thermodynamic equilibrium states are strongly affected, and new rules appear to govern the nature and the temperature of phase transitions.[4,14,15,16,17]

The understanding of the structure and phase behavior of LC's in confined geometry usually requires that finite size and surface anchoring effects are considered. The new boundary conditions imposed to the confined LC by the solid porous matrix can also be expressed in terms of an external field. Indeed, strongly disordered porous materials (such as aerogels) have been used in order to introduce random fields, which couples to the LC order parameters.[4,17] Under these conditions of confinement, the weakly first order isotropic-nematic and the continuous nematic-smectic phase transitions are shown to be mostly affected by the quenched disorder effects introduced by the random porous materials. This makes LC's confined in random porous matrices exceptional models to address questions of statistical mechanics that challenge current theories.[18]

The dynamics of LC's is complex, including several relaxation modes that occur at very different time and spatial scales. Accordingly, the dynamical behavior of confined LC's is very rich. Quasielastic light scattering (QELS) allows one to study the long time decay of the nematic fluctuations. On approaching the isotropic-nematic transition, this dynamics is related to the Laudau-de Gennes process corresponding to the randomization of the local nematic order parameter. It is characterized by a unique relaxation time in bulk, whereas the relaxation function is non-Debye in confined geometry and obeys a $\ln\tau$ scaling. In addition, the temperature dependence of the relaxation time is non-Arrhenius, which is a characteristic feature of glassy dynamics.[19] Both pore-pore interactions similar to spin glasses and random field effects have been invoked.[20] Such glassy behavior has also been observed by QELS in 8CB confined in aerogels, and has been associated to collective fluctuations of the smectic order parameter, in agreement with theoretical predictions based on quenched disorder effects.[4] Although it seems a quite general behavior, the connection between the glass-like dynamics of confined LC's and the nature of the porous topology, surface interactions and confinement-induced quenched disorder effects remains challenging.[21,22,23,24]

The individual molecular dynamics of confined LC's is also strongly modified. Dielectric relaxation spectroscopy has been extensively used since it allows one to probe the rotational-librational dynamics of the dipolar molecules around their short axis as well as tumbling relaxation modes around the local nematic director.[25] They correspond to fast relaxation modes, which are typically in the range of a few Mhz. The effect of confinement is described in terms of a glass-like dynamics too.[26,27] It includes the appearance of a broad distribution of relaxation times and a Vogel-Tamman-Fulcher (VTF) temperature dependence of the mean relaxation time. The slowing down on cooling of the orientational dynamics of confined LC's displays the heterogeneous and cooperative characteristics of glass-forming systems. Additionally, a slow process is observed and has been related to the hindered rotation of interfacial molecules.[28]

The ultra-fast rotational dynamics (from 1 ps to 10 ns) can be probed with the latest OKE techniques. These experiments have highlighted strong similarities between LC's and supercooled liquids on this short timescale.[29,30] In particular, scaling laws similar to the predictions of the Mode Coupling Theory (MCT) for glass-forming liquids are found.

On this particular timescale, many experimental studies have tracked the dynamics of glass-forming liquids under confinement. Among them, quasi-elastic neutron scattering experiments (QENS), have been able to probe together the vibrational, translational and rotational molecular dynamics.[31] They provided valuable information about the effects of surface boundary conditions, which are conveyed within the fluid to the inner pore and that may coexist with putative finite-size effects.[6,12] The occurrence of a slower and extremely heterogeneous dynamics in a time range covering the prediffusive to the main structural relaxation regimes is considered as a general feature of the influence from an attractive rigid wall.[32]

Conversely, QENS studies concerning the fast dynamics of confined LC's are scarce in the literature. In the present paper, we have studied the local short time dynamics of the model LC 4-n-octyl-4-cyanobiphenyl (8CB) by incoherent QENS. We have compared bulk 8CB to 8CB confined in porous silicon (PSi). The columnar form of PSi provides a model nanoporous geometry for confinement studies, which corresponds to straight one-dimensional channels.[33] This situation presents some similarity with previous studies of molecular liquids confined in porous materials. It allows for a comparison between nanoconfined normal (or supercooled) liquids and nanoconfined LC's in their isotropic phase.[6] In addition, we can consider additional features that occur specifically in LC's at lower temperature and are related to mesomorphic properties. We have recently studied the thermodynamical and structural behavior of 8CB in PSi by neutron diffraction.[34] The smectic transition is completely suppressed, leading to the extension of a short-range ordered smectic phase. It evolves reversibly over an extended temperature range, down to 50 K below the N-SmA transition of bulk 8CB. In the present paper, the molecular dynamics related to this exceptional behavior has been tracked as a function of temperature.

We first describe samples and experimental setups. Then, we present confocal micro Raman spectroscopy experiments, which provide a spatially resolved analysis of the 8CB loading in porous silica. An overview of phase transitions and the occurrence of a short-range ordered smectic phase[34] is described by coherent neutron diffraction experiments. Finally, we report new results obtained by incoherent neutron scattering, which probes the averaged atomic autocorrelation function. The individual molecular dynamics of bulk and confined 8CB is then described in terms of incoherent elastic intensity and mean square displacement. Temperature scans at fixed energy window (FWS) are presented, covering a wide temperature range that extends from the isotropic liquid down to mesomorphic and crystalline phases. Two complementary spectral ranges have been covered by neutron time of flight (TOF) and backscattering (BS) techniques.

## 2. Experimental procedure

### 2.1. Sample

Fully hydrogenated 4-n-octyl-4-cyanobiphenyl (8CB) was purchased from Aldrich and used without further purification. Bulk 8CB undergoes with increasing temperature the following sequence of phases: crystal (K), Smectic A (A), Nematic (N) and isotropic (I) with the following transition temperatures: $T_{KA}$=294.4 K, $T_{NA}$=305.8 K and $T_{NI}$=313.5 K.

Porous silicon matrices were made from a crystalline silicon substrate using an electrochemical anodization process in a HF electrolyte solution. Anodization conditions control properties such as porosity, thickness and pore size. Anodization of heavily p- doped (100) oriented silicon leads to highly anisotropic pores running perpendicular to the surface wafer (called columnar form of the PSi). These samples were electrochemically etched with a current density of 50 mA.cm$^{-2}$ in a solution composed of HF, H$_2$O and ethanol (2:1:2) according to previous studies.[35] These controlled conditions give a parallel arrangement of not-connected channels (diameter: ~300 Å, length: 30 μm) as shown in Fig. 1. The aspect ratio of each channel exceeds 1000:1 and confers a low dimensionality (quasi 1D) to the system. The preferential alignment of all the channels perpendicularly to the silicon surface prevents powder average limitations when measuring anisotropic observables of unidimensional nanoconfined systems. PSi is therefore a model porous material for studying nanoconfinement effects in a quasi-1D geometry with macroscopic order along the pore axis.

The inner surface of PSi was covered by a native oxide layer so that the interaction energy with 8CB is similar to other porous silica matrices. A complete thermal oxidation of the porous layer can also lead to a transparent matrix, which can be used for optical applications or characterizations.[36] An additional feature of PSi, which is at variance to other membranes such as anopores, is its strongly irregular inner surface at the nanometer scale. The latter has been recently shown to introduce strong disorder effects, which influence phase transitions such as capillary condensation or nematic to smectic transition.[34,37] PSi provides a unique opportunity to study quenched disorder effects with an anisotropic macroscopically aligned porous solid.

8CB was confined into PSi samples by impregnation from the liquid phase. The complete loading was achieved under 8CB vapour pressure in a vacuum chamber and at a temperature of 60°C, well above the N-I transition temperature. The excess of liquid crystal was removed by wiping the samples with Whatman filtration papers.

*2.2. Raman Spectroscopy*

Raman spectra were recorded with a HR800 Jobin-Yvon spectrometer using the 632.8 nm radiation of an He-Ne ion laser in a backscattering geometry at room temperature. Raman spectra of 8CB confined in fully oxidized PSi were collected in confocal geometry under an optical microscope. The use of an objective of x100 magnification allows to reach a spatial resolution of about 1.5 μm. Scans across the porous layer thickness were performed with a microtranslation stage.

*2.3. Neutron Diffraction*

The neutron scattering experiments were performed on the double axis spectrometer G6.1 of the laboratory Léon Brillouin (LLB) neutron source facility (CEA-CNRS, Saclay) using a monochromatic incident wavelength of 4.74 Å. A cryostat was used to control the temperature of the sample with a stability better than 0.1 K in a range from 180 K to 310 K. Eight PSi wafers with 30 μm porous layers each were stacked vertically in order to improve the signal/noise ratio. A cylindrical aluminium cell was used in order to change the sample orientation by a simple rotation of the cell without any significant variation of the corrections for sample holder scattering and screening. Two particular orientations corresponding to grazing and normal incidence were selected so that the transfer of momentum **q** for small $q$ values is parallel and perpendicular to the pores axis, respectively.

*2.4. Quasi Elastic Neutron Scattering*

Quasi Elastic Neutron Scattering (QENS) experiments were carried out with fully hydrogenated samples using the high resolution back scattering spectrometer (BS) IN16 at the Institut Laue Langevin (Grenoble) and the time of flight spectrometer G6.2 (TOF) at the LLB.[38] The contribution from the incoherent cross section of the hydrogen atoms correspond to 89% of the total scattering cross section of one 8CB molecule, so that the contribution to the measured scattering intensity from other atoms and from coherent scattering can be neglected within a good approximation. A standard configuration of the IN16 spectrometer was chosen with Si(111) monochromator and analysers in backscattering geometry, which corresponds to an incident wavelength of 6.271Å and results in an energy resolution (FWHM) of 0.9 μeV. The energy range was 15 μeV with a $q$ range between 0.2 Å$^{-1}$ and 1.9 Å$^{-1}$. The time of flight G6.2 spectrometer allows one to extend the energy transfer range with a relaxed resolution (FWHM) of 107 μeV at the elastic position for an incident wavelength of 6 Å. The elastic $q$ range covered by G6.2 in this configuration extends from 0.45 Å$^{-1}$ to 1.95 Å$^{-1}$.

PSi layers were placed in cylindrical aluminium cells similarly to diffraction experiments. A cryofurnace and a cryoloop, respectively, were used on IN16 and G6.2 spectrometers in order to regulate the sample temperature in a range from 10 K to 340K or 100K to 340K, respectively.

Different orientations of the pore axis with respect to momentum transfer vector **q** (grazing incidence, 45°, 135° and normal incidence) were obtained by a simple rotation of the cell. On IN16, grazing and normal incidences with respect to the surface of the PSi-wafers respectively correspond to **q** vectors being almost parallel and perpendicular to the pore axis (only for the detectors selecting the smallest transfers of momentum, i.e. of the order of 0.2 Å$^{-1}$). These two scattering geometries could not be chosen on G6.2 since the smallest reachable scattering angle is 2θ = 25.5 °. Incidences angles of 45° and 135° were then used on both spectrometers, for which **q** vectors at 1.48 Å$^{-1}$ (1.42 Å$^{-1}$) on IN16 (G6.2), are oriented perpendicular (parallel) to the pore axis.

Fixed window scans (FWS, i.e. with Doppler device stopped) were carried out on IN16 in a temperature range from 10 K to 340 K. FWS were also recorded on G6.2 with an energy resolution 120 times larger than IN16, on a temperature range from 100 K to 340 K, by summing the time-of-flight channels in the elastic peak region.

Standard data corrections were applied using conventional programs provided at ILL (sqw) and LLB (quensh). The incoherent scattering function $S_{inc}(q,\omega)$ is commonly approximated by Eq. 1 :

$$S_{inc}(q,\omega) = \exp\left(-\frac{\langle u^2 \rangle q^2}{3}\right)\left[A(q)\delta(\omega) + (1-A(q))S_{quasi}(q,\omega)\right] \quad (1)$$

The first factor is the Debye-Waller term that reflects vibrational motions.[39] The second factor includes the elastic incoherent structure factor $A(q)$ (EISF) and the quasielastic component due to relaxing processes. At low temperature, the quasielastic lines are not broad enough to be discriminated from a true elastic peak within the experimental resolution. In this temperature range, the $q^2$-dependence of the logarithm of the elastic intensity allows one to measure the mean square displacement (MSD) $\langle u^2 \rangle$ of vibrational modes still populated at low temperature. In the harmonic approximation of a solid, $\langle u^2 \rangle$ should increase linearly with $T$.

Above a certain temperature, when the molecular relaxation processes become faster than the corresponding energy resolution, this harmonic approximation breaks down. The quasielastic broadening that exceeds the fixed energy window does not contribute to the measured elastic intensity any more. This onset of relaxational contributions can be easily detected by FWS. In this temperature range, the quasielastic scattering contribution reduces the elastic scattering and adds thus to the truly vibrational part, resulting in an increased effective MSD $<u^2_{eff}>$.

The Debye-Waller factors have been extracted from the FWS by a linear regression of the logarithm of the elastic intensity as a function of $q^2$ on a $q$-range from 0.7 Å$^{-1}$ to 1.9 Å$^{-1}$. The scattering intensity at smaller $q$-values has been discarded in the fitting procedure in order to avoid undesired contributions from the increasing coherent scattering at low $q$ in the region of the smectic Bragg peak, and from the mesostructured composite material.

FWS have been performed both on cooling and heating with typical ramps from 0.5 K.min$^{-1}$ to 2 K.min$^{-1}$. Data accumulation was performed during the continuous temperature scan integrating over intervals of 1 to 2 K.

3. Results

*3.1. Raman Characterization of the 8CB loaded in the channels*

Micro Raman spectroscopy was used to check that PSi channels were homogeneously filled by 8CB after the loading procedure described in part 2.1. For this specific experiment, a complete thermal oxidation of the porous layer was performed before confinement. It leads to an optically transparent porous layer, which significantly improves the experimental conditions for a Raman characterization. Although it is known that a complete oxidation of PSi quantitatively changes its morphological parameters (porosity, pore diameter), a thin oxide layer covers always the inner pore surface and it does neither affect significantly its topology nor the interfacial interaction energy. .

Raman spectroscopy allows one to identify lines, which are characteristic of intramolecular vibrational modes of 8CB.[40] The spatial resolution (FWHM) of the confocal micro Raman spectroscopy is about 1.5 μm (calibrated on a silicon plate substrate). A direct scan and check of the 8CB loaded in PSi across the porous layer is obtained by changing the location of the volume analyzed by the confocal microscope device. Fig. 2 displays several spectra obtained for different positions of 8CB in the porous layer (thickness of 10 μm as measured by SEM) between 1400 to 2300 cm$^{-1}$ at room temperature. The strongest lines at 1609 and 2220 cm$^{-1}$ are associated to aromatic C-C and C≡N stretching modes of 8CB, respectively. The intensity profile of these two lines clearly exhibit a first steep increase when the focused spot enters the porous layer. Then it remains almost constant across the porous layer and finally sharply vanishes. An additional line appears concomitantly at 520 cm$^{-1}$. It is attributed to a Si crystal mode and proves that the volume checked at this point corresponds precisely to the bottom of the oxidized channels. This demonstrates that 8CB is homogeneously present in the entire porous layer. In addition, it shows sharp interfaces at the air/PSi and PSi/Si substrate boundaries.

*3.2. Phase behavior by diffraction*

The static structure factors obtained by neutron diffraction for bulk 8CB and under confinement in PSi are displayed in Fig. 3 for various temperatures corresponding to the four different stable phases of the bulk. The spectra are flat at 315 K and 310 K in bulk and PSi, but in spite of the large incoherent scattering cross section of the hydrogen atoms, which tends to mask the weak diffuse scattering, a typical short-range order peak can be observed in the isotropic and nematic phases near 0.2Å$^{-1}$.

This is the *q* value at which at lower temperature, bulk 8CB shows a single peak when transforming to a Smectic A phase.[41] Bulk 8CB eventually transforms to a crystalline phase at 294 K. Note that the spectrum acquired at this temperature clearly shows a coexistence of crystalline phases during the transformation with a small peak at 0.2 Å$^{-1}$ along with additional Bragg peaks. The stable crystalline phase that shows up at this temperature has been described by Kuribayashi et al..[42] It is characterized by an intense peak at about 0.48 Å$^{-1}$ and other peaks at 1.15 Å$^{-1}$ and 1.4 Å$^{-1}$.

8CB confined in PSi exhibits a very different behavior at temperatures below the bulk nematic-to-smectic transition. Indeed, no crystallization is observed down to 250 K and a single peak at 0.2 Å$^{-1}$ reflecting a smectic order is observed. The temperature variation of the maximum intensity of this peak is reported in Fig. 4, which depicts the phase diagram of bulk and confined 8CB. It shows that confinement strongly depresses the crystallization temperature. A complete analysis of the crystallization process that occurs at lower temperature has been performed from diffraction experiments and is not in the focus of the present contribution.[43] It reveals the formation of two different crystalline phases on cooling that are reflected by a two-step decrease of the smectic Bragg peak intensity in Fig. 4 at about 250K and 230K. Simultaneously, crystalline Bragg peaks corresponding to two different phases appear successively for *q*-values of 0.25 Å$^{-1}$ and 0.15 Å$^{-1}$, as shown in dashed lines in Fig. 4. These phases, which do not correspond to stable crystalline phases of bulk 8CB have been reported in confined geometry.[40,44] In the temperature region covered on cooling from 250 K to 220 K, three phases of confined 8CB coexist: two crystalline polymorphs and the short-range smectic phase.

Most noticeable for our present interest is the drastic change of the nematic-smectic transition at higher temperature. Indeed the sharp phase transition that occurs at 305.8 K for the bulk is replaced by the gradual growth of a smectic order on cooling on an extended temperature range from 305 K to 250 K for the confined system. A consistent analysis of the shape of the Bragg peak with a functional form predicted by recent random field theories has been applied (cf. Ref. 34 for technical details). In particular, it shows that the smectic order is restricted to short-range with an associated longitudinal correlation length growing reversibly from 4 to 15 nm on the temperature range from 305 K to 250 K. The occurrence of this short-range smectic phase, lately denoted as SRS in the text, can only to a negligible extent be related to surface energy and finite size effects introduced by confinement. This latter situation is conversely encountered when 8CB is confined in 1D-channels with smooth surfaces formed by anopore (alumina) or nuclepore membranes.[13,45] It leads to a moderate broadening of the nematic-smectic transition and a small decrease of the transition temperature (typically 2 K). These two phenomena mostly reflect cut-off of the critical divergence of the smectic correlation length due to the geometrical confinement and a temperature depression due to the adding of elastic energy and an interfacial energy term to the volume free energy of the confined phases.

The occurrence of the SRS phase of 8CB in PSi illustrates another effect introduced by mesoporous confinement, which is the quenched disorder effect. Quenched disorder in strongly disordered porous materials arises from the random interaction between the matrix and the confined LC, which acts as random fields coupling to the nematic and smectic order parameters. Such a situation is mostly encountered for random porous materials (aerogels and aerosils dispersions), which are isotropic and homogeneous materials.[4,15]

PSi geometry displays a porous morphology formed by straight 1D-channels with strongly irregular pore surfaces at the nanometer scale. Consequently, adjacent confined molecules may experience very different surface interactions (preferential pining), which frustrates the growth of smectic correlations within the pore. The balance between the strength of quenched disorder and the smectic elastic energy when the temperature decreases leads to the progressive growth of a saturating smectic correlation length.[18] A more detailed analysis of the shape of the Bragg peak shows that the increase of the smectic correlations on cooling firstly arises from thermal fluctuations, while the onset of a contribution due to static disorder dominates at lower temperatures (below 295 K).[34]

This observation is consistent with recent theoretical predictions about the smectic behavior in the presence of random fields, although available models have been developed in the limit of weak disorder, which is not likely to be the case encountered for PSi.[4,18] PSi allows to reach an unexplored regime of strong disorder with respect to more studied random porous materials (aerogels and aerosils dispersions)[4,15] In addition, the one-dimensional topology of the PSi channels confers a novel anisotropic character to this quenched disorder. As a consequence, the SRS is not isotropic, as it is usually in random porous materials, but develops smectic correlations preferentially in the direction of the nanochannels long axis. This is demonstrated by the existence of a Bragg peak at $q \approx 0.2$ Å$^{-1}$ in grazing incidence only with respect to the PSi wafers surface. A preferential orientation of the molecules along the pore axis in the same temperature range has been observed by spectroscopic ellipsometry too.[46]

### 3.3. Molecular dynamics

Fixed window scans have been performed using the two complementary spectral resolutions of BS- and TOF-spectrometers (probing roughly ns- and ps-time scales). They have been acquired during temperature scans, with corresponding cooling or heating rates of the order of 0.5 K.min$^{-1}$ or 2 K.min$^{-1}$ and integrating typically over 2 K.

The molecular dynamics of 8CB confined in PSi as probed by FWS on both instruments presents no significant anisotropic character, despite the preferential orientational and smectic translational orders along the pore axis. Indeed, the elastic intensity is weakly affected by a variation of the orientation of the sample with respect to the incident neutron beam. This is illustrated in Fig. 5 by the elastic window intensity measured by BS at T=280 K as a function of $q^2$ for four different angles of incidence. These intensities obtained for different orientations of **q** with respect to the pore axis are reasonably comparable within the experimental uncertainties. The logarithm of the elastic intensity acquired for the various angles of incidence shows a same linear $q^2$-dependence, which allows one to extract an effective isotropic mean-square displacement $<u^2>$ as discussed in section 2.4. The absence of any strong anisotropic character in the BS- and TOF-neutron scattering has to be related to the large $q$ values and short timescales covered by these techniques. They essentially probe local molecular dynamics at spatial and time scales where the anisotropy of the system does not show up. This is probably not the case anymore for $q$ values of the order of the inverse pore size or for long time relaxation processes, such as large scale translational diffusion and for collective or order parameter relaxation processes. This conclusion reached for a confined nematogen agrees with a recent study of liquid alkane chains confined in PSi.[47]

Fig. 6 displays the elastic intensities as a function of temperature after integration over the full scattering angles in order to improve the signal quality. They have been corrected for the contribution arising from the empty cell and the empty Si wafers and normalized to the lowest temperature intensity. These background contributions were in the range from 10% to 20% of the maximum intensity at a temperature of 300K and 100K respectively. The corrected elastic scattering reflects besides possible minor coherent contributions the purely elastic components of an hypothetical scattering function (elastic incoherent structure factor (EISF) from restricted motions), the part of the relaxation processes that occurs at low frequency within the spectrometer resolution (or very low frequency inelastic scattering), weakened by the Debye-Waller factor. For the bulk liquid above 300K, the elastic intensity is very small on IN16 (about 5% of its maximum value), as expected for a liquid for which translational diffusion completes within the experimental timescale. The convolution with the translational diffusion quasi-elastic line broadens the hypothetical purely elastic component (EISF) of spatially localized relaxation processes (tumbling, rotational degrees of freedom). The 120 times broader spectral resolution of TOF as compared to BS leads to a larger elastic intensity at the same temperature (about 45%). The difference of resolution alone cannot account for the difference of measured elastic scattering without invoking different relaxation processes. Indeed, if one assumes that the elastic scattering probed by the two spectrometers arises from an unique relaxation process (single Lorentzian quasi-elastic line), one would expect to measure a larger elastic intensity by TOF (of the order of 90 % at 340K). The lower elastic scattering measured by TOF indicates that additional quasielastic fast relaxation processes occur in the frequency window of TOF in the liquid phase but are too broad to contribute significantly in the BS quasielastic window (they appear in BS experiments as a weak flat background). The existence of different relaxation processes (translational diffusion and faster molecular rotation, tumbling and chains motion) are therefore expected to contribute with different strength to the scattering intensity in the different frequency ranges covered by BS and TOF.[25,26]

The elastic intensity of bulk 8CB progressively increases on cooling from the liquid state down to a temperature of about 275K. A sharp jump of the intensity occurs at this temperature, while it keeps increasing moderately on further cooling down to the lowest temperature measured. This discontinuity is assigned to crystallisation. Approaching the crystallization temperature from above, the probed molecular relaxation processes slow down progressively with temperature. However, this dynamics is not strongly affected by the liquid-crystal ordering since no signature of the isotropic-nematic nor the nematic-smectic transition is observed. This is at variance with crystallization, which leads to a sharp increase of the elastic intensity to 70% of its maximum, corresponding to a freezing of most of the relaxation processes. In particular, the absence of translational diffusion due to the localisation of the molecules at the sites of a crystalline lattice leads to a finite EISF. Some remaining degrees of freedom (methyl libration/rotation and Debye-Waller factor, comprising inter- and intramolecular vibrational modes) result in a remaining (but weaker) temperature dependence of the elastic component in the crystal phase, down to temperatures below 100 K where the vibrational contribution becomes dominant. Selective isotopic labelling of the alkyl chain should be a way to assess these side-chain contributions in future experiments.

The elastic intensity measured in confined geometry on cooling and heating are displayed in filled and open circles, respectively. It should be first noticed that the remaining elastic component is large even well above the clarification point (20% at 315 K on the longest time scale). It means that confinement introduces some either much restrained or much slower relaxation processes in the liquid phase. This is most probably a consequence of the dynamical boundary conditions introduced by confinement. Surface effects are known to strongly reduce the molecular mobility of the interfacial liquid.[6,12,31,32] Our observation complements similar reports on the case of simpler molecular liquids. Surface hindered dynamics has also been reported by dielectric spectroscopy for 8CB confined in porous matrices.[25,27,28]

On decreasing temperature, the elastic intensity measured by BS progressively increases and does not present any sharp transition. Eventually the low temperature branch reaches the level of the bulk elastic intensity at 220 K, which suggests that full crystallization is only achieved near this temperature. Clearly, we observe molecular dynamics that reflects different phase behaviors of bulk and confined 8CB. We have shown that 8CB confined in PSi progressively transforms into a short-range ordered smectic phase when cooling down from 305 to 250 K, with strongly depressed crystallization as shown in Fig. 4. The temperature variation of the elastic intensity measured by BS clearly demonstrates that this unusual phase behavior goes along with a continuous slowing down of the molecular dynamics (cf. Fig. 6(a)). A similar glass-like behavior has been observed in confined molecular liquids on cooling and is attributed to the propagation of the dynamical boundary conditions towards the centre of the confined system due to a dynamical correlation length.[6,12,32] However, these effects could only be observed in supercooled liquids confined within smaller pores, due to the very limited size of the dynamically cooperating regions in glass-forming systems.[5,7] The existence of larger correlations in 8CB and especially the gradual growth of a smectic correlation length on cooling within the SRS phase could be a key feature at the origin to this glass-like slowing down in the much larger pores of PSi.

Crystallisation of the confined phase is gradual and occurs in a temperature range from 250 K to 220 K, where two polymorphic crystalline phases are formed, as discussed in Part. 3.2. The continuous increase of the elastic intensity shown in Fig. 6(a) for the same temperature range is attributable to the varying fraction of coexisting SRS and polymorphic solid phases. An almost reversible behavior is obtained during the heating scan shown as open circles in Fig. 6(a) apart from the temperature range from 220 to 260 K, where a small hysteresis is observed. It may be assigned to the expected hysteresis loop of the melting-freezing transition.

The worse spectral resolution of TOF does not allow one to catch the entire slowing down of the molecular dynamics of the SRS phase on cooling down to 250 K as shown in Fig 6(b). In the SRS phase, the relaxation processes that lead to a quasielastic broadening and a decrease of the elastic part of the scattering intensity in the energy range of BS are essentially included within the elastic peak of TOF. Only the fastest part of the distribution of relaxation processes of 8CB, which are probably associated to molecular tumbling and chain torsional motion are observed by TOF. On decreasing temperature, the elastic intensity of the SRS reaches the level of the bulk crystal at 285K, which is the temperature at which large amplitude relaxation processes in the SRS phase occur on timescales longer than the TOF resolution. Below 285K, the SRS looks like a glassy phase, with remaining methyl group rotation and vibrational modes whereas additional slower relaxation processes are still observed by BS down to the crystallization temperature.

The high frequency vibrational modes that extend beyond the quasielastic energy window essentially contribute in terms of a Debye-Waller factor in the elastic range. The temperature and $q$-dependence of the elastic intensity obtained by FWS can in a first approximation be treated as an effective Debye-Waller factor, at least at sufficiently low temperature. This means that the logarithm of the elastic intensity should exhibit a linear $q^2$-dependence. Using this harmonic approximation allows one to estimate an average atomic mean squared displacement (MSD) $<u^2>$ as shown in Fig. 5. At higher temperature, the data are expected to depart from this linear behavior. Indeed, such deviation is observed at low $q$ ($q$ lower than 0.7Å$^{-1}$), but contributions other than the $q^2$-dependence of the elastic intensity are also possible in that range, such as coherent scattering form 8CB or also the huge small angle elastic coherent scattering associated to the mesoporous structure of the material. Therefore, this part of the spectra has been discarded in the fitting procedure (cf. dashed line in Fig. 5, and section 2.3).

The temperature variation of $<u^2>$ extracted from BS measurements for bulk 8CB displays three distinct regimes (cf. Fig. 7). At high temperature, large values of the MSD of the order of 2 Å$^2$ are observed in the smectic phase (not shown). In this fluid phase, the harmonic approximation breaks down and the effective values of $<u^2>$ include relaxational contributions (translational diffusion and rotation) in terms of quasielastic components and EISF. On cooling, a sharp drop of $<u^2>$ from about 2 Å$^2$ to values of the order of 0.3 Å$^2$ occurs at 275 K. It provides a clear signature of crystallisation during the cooling scan. Additional intramolecular fast relaxation processes (mainly methyl rotation but also end chain motion), contribute to the effective MSD in the crystalline phase down to 100K. Below 100 K in the crystalline phase, a linear increase of the MSD with temperature is observed. This feature is fully consistent with the Debye harmonic approximation of a solid phase.

The values of the MSD obtained by TOF present a similar temperature variation with some quantitative differences. The latter values are systematically smaller by a factor of two than the one obtained by BS. This feature is known and attributable to the different energy resolution between the two techniques, which leads to a different contribution from quasielatic scattering and EISF to the effective mean squared displacement in the temperature range where the harmonic approximation is not strictly valid anymore.

The MSD of 8CB confined in PSi obtained by BS on cooling present many differences from the bulk one (cf. circles in Fig. 7(a)). It shows a plateau at about 0.9 $\text{Å}^2$ in a region above 250 K, which corresponds to the confined SRS phase. This value is significantly smaller than the one obtained in the bulk smectic phase. The effective MSD obtained in this temperature range are actually very sensitive to differences in the molecular relaxation processes that may contribute in the energy range covered by BS in terms of quasielastic scattering and EISF. Translational diffusion relaxation processes are observed in the bulk smectic phase in the $\mu$eV energy range. The low value of the MSD in the confined 8CB indicates that the molecular dynamics of the SRS in the same energy range comprise motions that are more restricted in space.

On cooling, this plateau extends to 250K, which corresponds to the region of stability of the SRS phase. The deviation from this plateau is smooth and does not compare with the sharp jump observed in the bulk at higher temperature, which has been attributed to a sudden crystallization. The MSD decreases continuously from 0.9$\text{Å}^2$ to 0.3$\text{Å}^2$ in the range from 250 to 220 K, where it roughly compares the MSD of the bulk crystalline phase. This reflects the freezing of the molecular motion during the gradual crystallization of the SRS phase, following the decrease of the integrated intensity of the smectic Bragg peak on the same temperature range as shown in Fig. 4. The measured MSD reflects an effective value averaged on the coexisting SRS and polymorphic solid phases, which have very different molecular mobility in the frequency range probed by BS.

Below 220K, the MSD of the confined solid phases is close to the bulk one. It reflects pure vibration contributions for temperatures below about 100K and additional fast relaxation processes in the range from 100 K to 220 K, which are attributable to methyl group motion. Relaxation processes in this intermediate region are presumably too slow compared to the TOF spectral resolution, which explains why this region is mostly overlooked in the TOF-experiments.

## 4. Conclusion

LC's in nanopores allow one to address various crucial aspects of the confinement of molecular fluids such as finite size, interfacial, low dimensionality and quenched disorder. They are likely to produce different effects on the structural, thermodynamical and dynamical properties of the confined phase depending on the system under investigation. We have shown that the structure of 8CB is sensitive to the anisotropy of confinement of PSi, leading to preferential directions for nematic and smectic ordering. The phase behavior is profoundly modified in confinement. 8CB progressively develops only short-range smectic correlations from ambient to remarkably low temperatures with no signature of a true phase transition. The existence of a SRS phase in confinement is interpreted as the consequence of strong quenched disorder introduced by PSi. The molecular dynamics of the confined liquid, nematic and SRS phases are also profoundly affected. A strong reduction of the molecular mobility is observed in the liquid phase at temperatures as high as 315 K, which is well above the nematic transition. About 20 percents of the incoherent intermediate scattering functions lays in the energy resolution of BS at this temperature. This is attributed to the dynamical boundary conditions introduced by confinement, which lead to a slower and restricted molecular motion. On decreasing temperature, a glass-like increase of the elastic intensity is observed in the temperature range of stability of the confined SRS phase. It is at variance to the bulk elastic intensity, which shows a discontinuous transition from a fluid smectic phase to a frozen crystal.

This glass-like slowing down of the molecular dynamics compares with the behavior of supercooled liquids confined in smaller pores (diameter less than 10nm), where the existence of a dynamical correlation length has been invoked as a way to propagate the reduced molecular mobility from the pore surface towards the centre of the confined liquid. In the latter case, the dynamical correlation length is limited to a few molecular diameters and there is no indication for any growing static correlation length.

The specificity of 8CB confined in PSi is the gradual growth on cooling of a short ranged order in the SRS phase, which satisfies the irregular surface interactions of PSi. The slowing down molecular mobility on cooling to 250K appears while a smectic local order develops in the pore. The present result makes 8CB confined in PSi a unique system for a detailed analysis of the influence of a growing nanometric static correlation length on the dynamics of a confined fluid. This gives the unique opportunity to analyse confinement effects on the slowing down of the structural relaxation of a complex liquid at a typical length scale about 10 times larger than the one attached to simpler liquids or glass formers. In addition to the transmission of the dynamical boundary conditions to the confined phase, a direct consequence of a strong quenched disorder and its interrelation with finite-size and low dimensionality effects have also to be considered in the short-range smectic phase.

**Acknowledgements**

We thank J. P. Ambroise and I. Mirebeau for the help in the experiments performed at the Laboratoire Léon Brillouin neutron source facility (France). The use of the Raman spectrometer from the Europia/ONIS platform at the University of Rennes and financial supports from the *Centre de Compétence C'Nano Nord-Ouest* and *Rennes Metropole* are expressly acknowledged.

[47] P. Huber, R. Zorn, B. Frick et al., Experimental Report, ILL #6-02-357 (2005).

**FIGURE CAPTIONS**

**FIG. 1:** Scanning electron micrographs of the porous silicon film. (a) side view at low magnification showing the 30 μm thick porous layer attached to the silicon substrate and (b) top view at higher magnification.

**FIG. 2:** MicroRaman spectra showing intramolecular vibrational bands (CN and CC stretching modes) of confined 8CB while scanning across the porous layer of thickness 10μm.

**FIG. 3:** Neutron diffraction structure factor at various temperatures of bulk 8CB (a) and 8CB confined in PSi (b) acquired under grazing incidence.

**FIG. 4:** Phase behavior of bulk and confined 8CB, revealed by the temperature variation of the integrated intensity of some selected Bragg peaks characteristic of the different phases. The smectic Bragg peak of bulk (open circle) and confined 8CB (filled circles) is located at $0.2\text{Å}^{-1}$. The crystalline Bragg peaks corresponding to two different phases appear successively for $q$-values of $0.25\text{ Å}^{-1}$ and $0.15\text{Å}^{-1}$ (squares and triangles respectively).

**FIG. 5:** (a) $q$-dependence of the elastic intensity measured by backscattering at T=280 K for 8CB confined in PSi and for different angles between the incident neutron beam and the Si wafers plane. Grazing and normal incidences correspond to $\omega$ values of 0° and 90° respectively.

**FIG. 6:** Elastic scattering of 8CB confined in PSi (circles) or bulk (solid line), measured for two different elastic window resolutions with (a) backscattering and (b) time-of-flight spectrometers. Both cooling and heating scans are shown for 8CB in PSi (corresponding respectively to filled and open circles). All other displayed curves have been obtained on cooling. The intensity is corrected for empty sample contribution, integrated from 0.4 to 1.9 Å$^{-1}$ and normalised at the lowest temperature.

**FIG. 7:** Mean square displacement of 8CB confined in PSi (circles) or bulk (solid line), on cooling with (a) backscattering and (b) time-of-flight spectrometers.

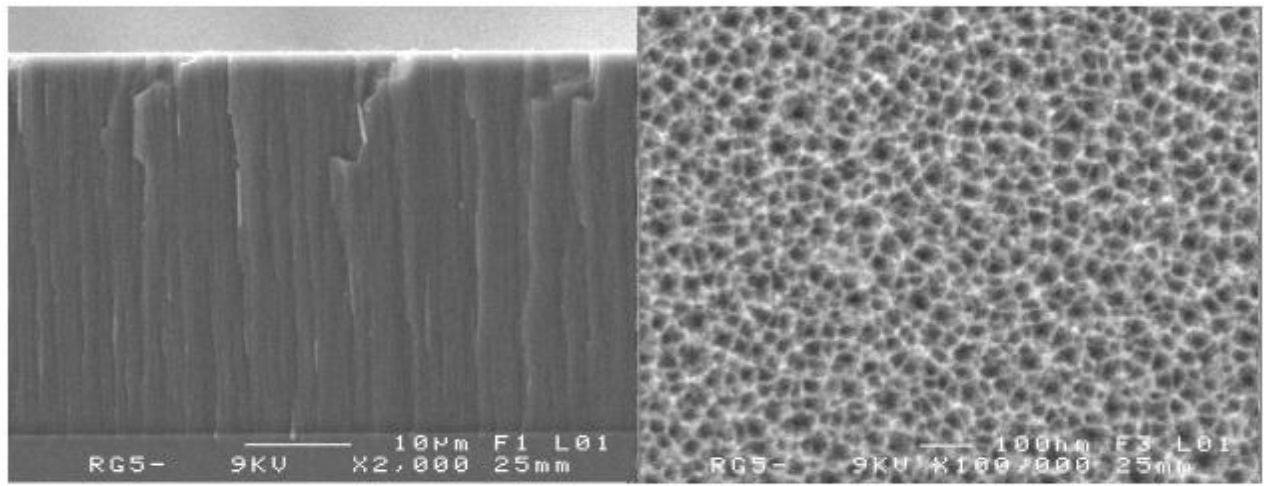

Figure 1
**Molecular dynamics of a short range ordered smectic phase nanoconfined into porous silicon.**
R. Guégan et al.

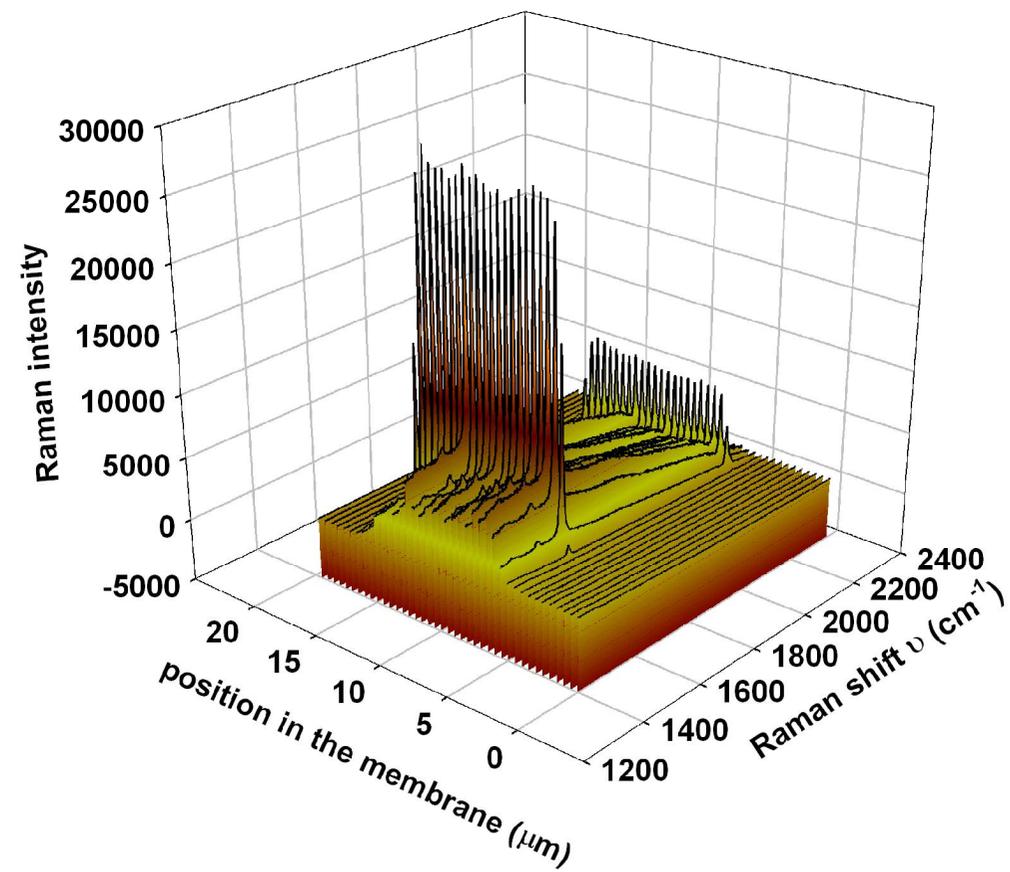

Figure 2
**Molecular dynamics of a short range ordered smectic phase nanoconfined into porous silicon.**
R. Guégan et al.

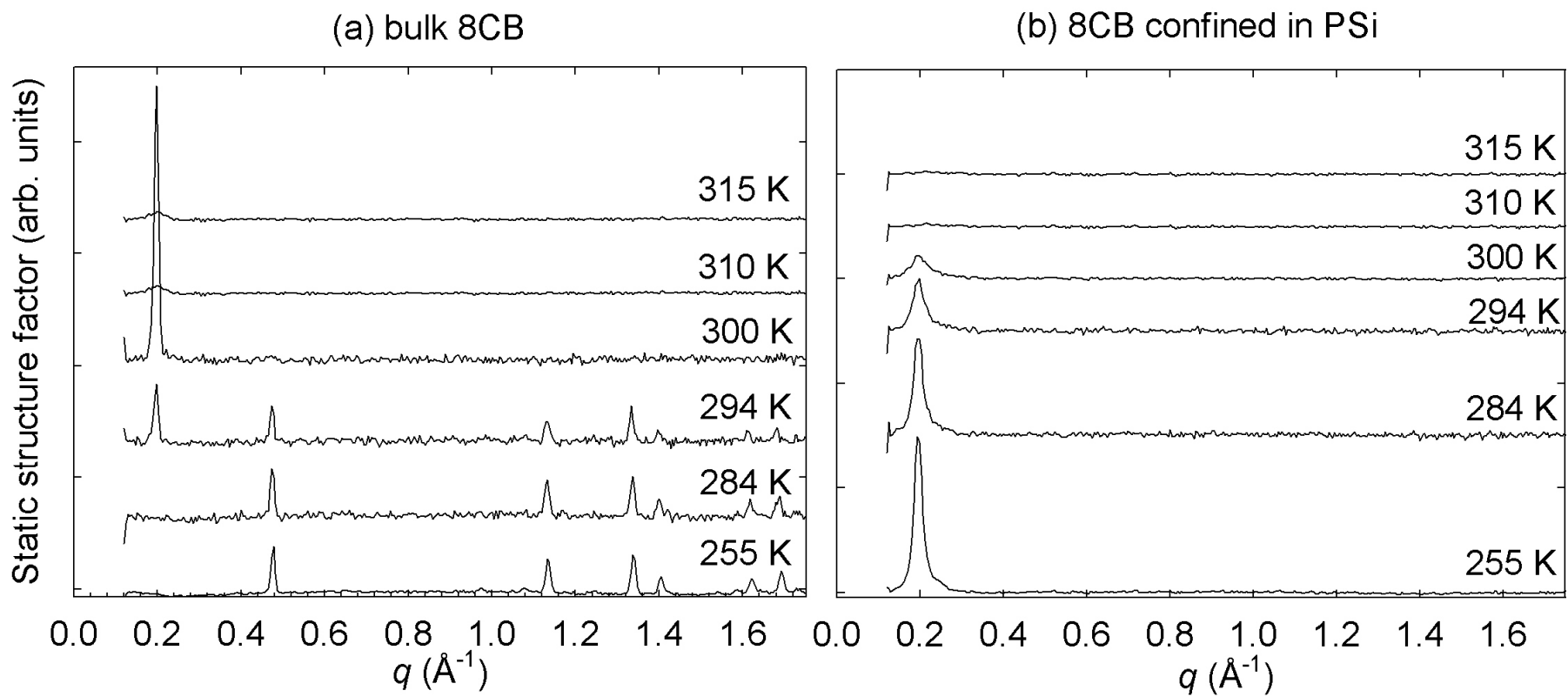

Figure 3
**Molecular dynamics of a short range ordered smectic phase nanoconfined into porous silicon.**
R. Guégan et al.

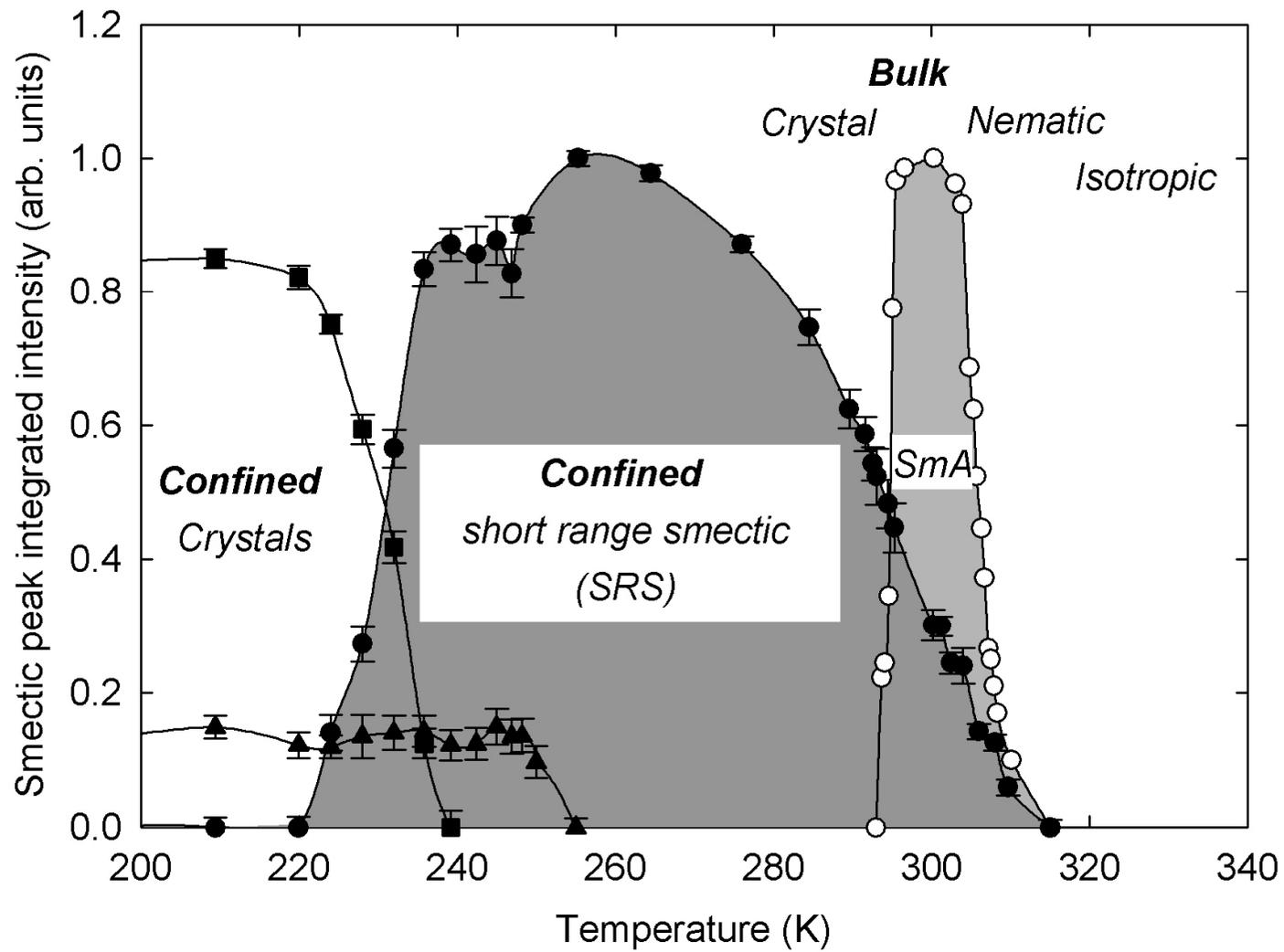

Figure 4
**Molecular dynamics of a short range ordered smectic phase nanoconfined into porous silicon.**
R. Guégan et al

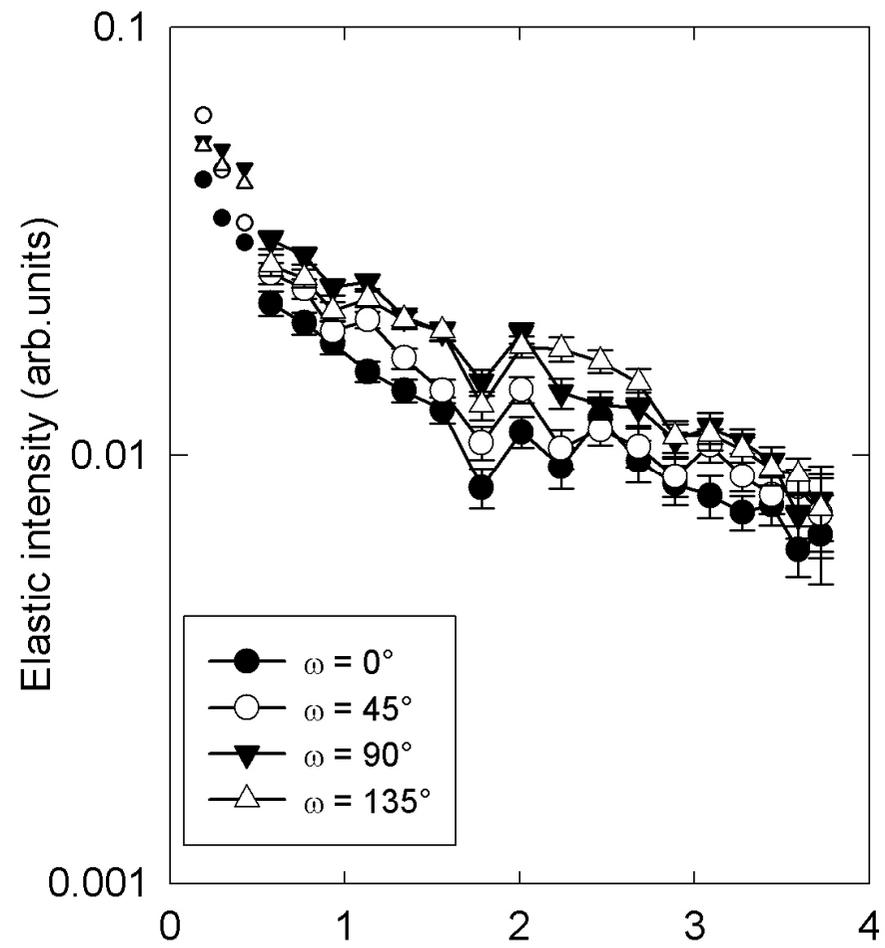

Figure 5
**Molecular dynamics of a short range ordered smectic phase nanoconfined into porous silicon.**
R. Guégan et al.

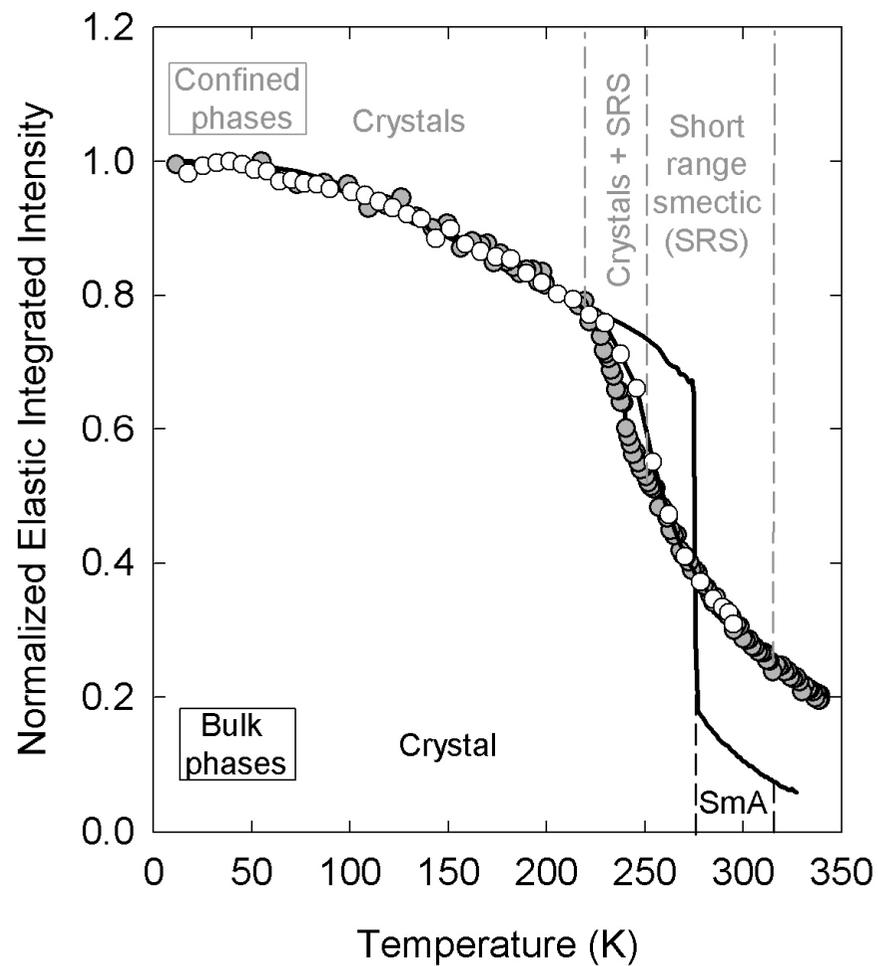 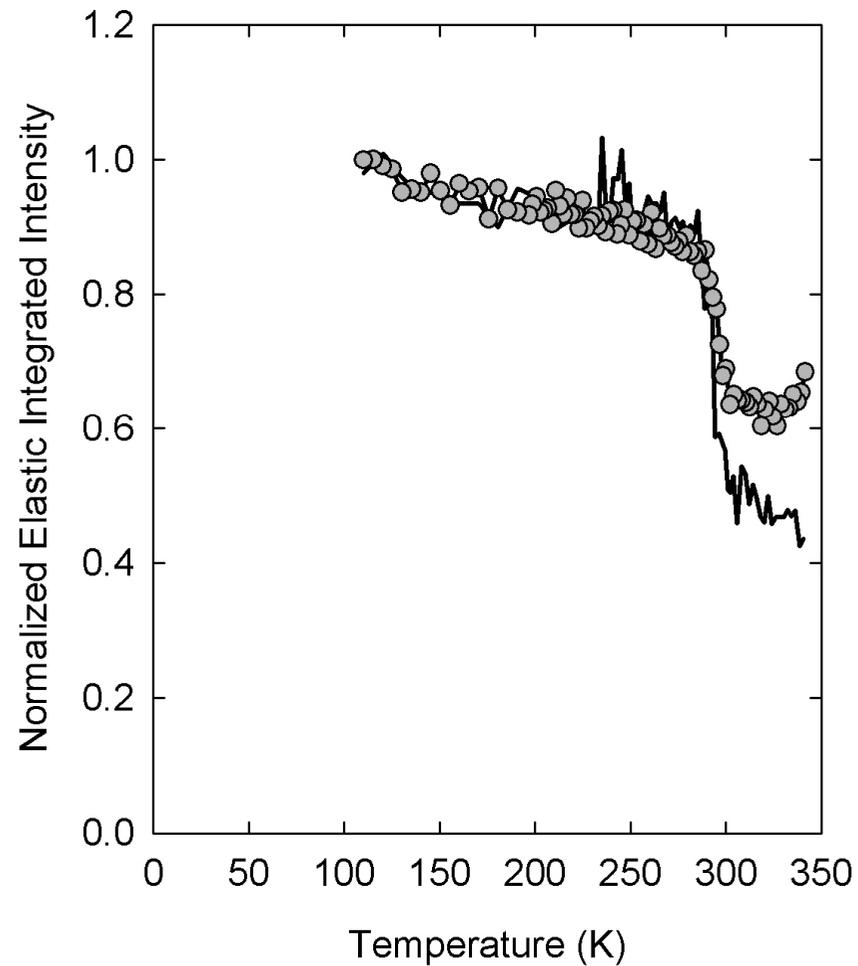

Figure 6
**Molecular dynamics of a short range ordered smectic phase nanoconfined into porous silicon.** R. Guégan et al.

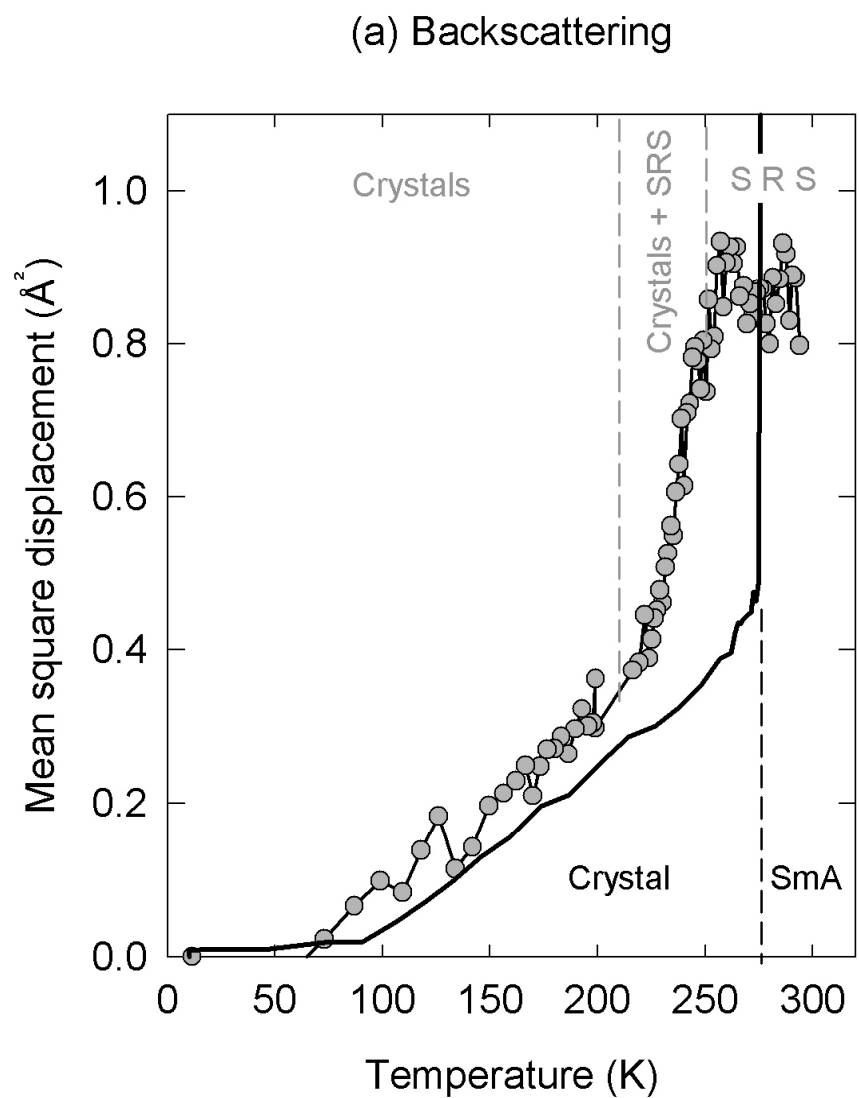 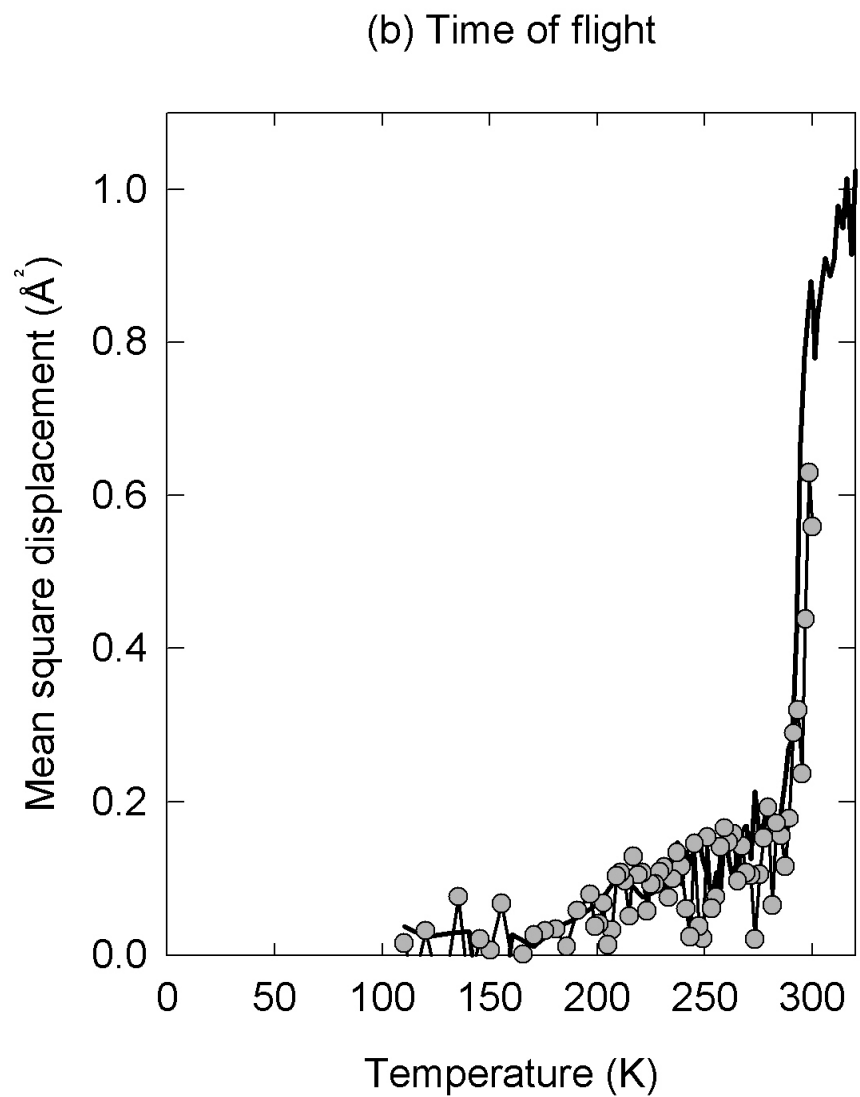

Figure 7
**Molecular dynamics of a short range ordered smectic phase nanoconfined into porous silicon.**
R. Guégan et al